\documentclass[a4paper]{jpconf}
\usepackage{graphicx}
\usepackage{cite}
\usepackage{color,amsmath,amssymb}
\begin{document}
\title{Thermodynamics of the frustrated
ferromagnetic spin-1/2 Heisenberg chain}

\author{J. Richter$^{1}$, M.~H\"artel$^1$, D.~Ihle$^2$ and
S.-L.~Drechsler$^3$}

\address{$^1$ Institut f\"{u}r Theoretische Physik,
Universit\"{a}t Magdeburg, D-39016 Magdeburg, Germany}

\address{$^2$ Institut f\"{u}r Theoretische Physik, Universit\"{a}t Leipzig,
D-04109 Leipzig, Germany}

\address{$^3$  Leibniz-Institut f\"{u}r Festk\"{o}rper- und
Werkstoffforschung Dresden, D-01171 Dresden, Germany}

\ead{johannes.richter@physik.uni-magdeburg.de}

\begin{abstract}
We  studied
the thermodynamics 
of the one-dimensional $J_1$-$J_2$ 
spin-{1/2} Heisenberg chain 
for ferromagnetic nearest-neighbor bonds $J_1 < 0$ and frustrating
antiferromagnetic next-nearest-neighbor bonds $J_2 > 0$
using  full diagonalization of finite rings
and a
second-order Green-function formalism. Thereby
we focus on $J_2 < |J_1|/4$ where the ground
state is still ferromagnetic, but the frustration influences the
thermodynamic properties.      
We found that 
their critical indices
are not changed 
by $J_2$. The analysis of the low-temperature
behavior of the 
susceptibility $\chi$
leads to the conclusion that 
this behavior changes from $\chi \propto T^{-2}$
at $J_2 < |J_1|/4$ to $\chi \propto T^{-3/2}$ 
at the quantum-critical point $J_2=|J_1|/4$.  
Another 
effect of the 
frustration is the
appearance of an extra low-$T$ maximum in the specific heat 
$C_{v}(T)$ for
$J_2 \gtrsim |J_1|/8$,
indicating its strong influence 
on the low-energy spectrum.
\end{abstract}

{\it Introduction: }
In low-dimensional frustrated quantum magnets 
 thermal and quantum fluctuations strongly influence the low-temperature
physics \cite{sachdev,book}.  
Special attention has been paid to one-dimensional (1D) $J_1$-$J_2$ quantum Heisenberg
magnets, see Ref.~\cite{Mikeska04} 
and references therein.
Recent experimental studies  have shown 
that  edge-shared chain
cuprates, such as LiVCuO$_4$, Li(Na)Cu$_2$O$_2$,
Li$_2$ZrCuO$_4$, and Li$_2$CuO$_2$
\cite{gibson,matsuda,gippius,ender,drechs1,drechs3,drechs4,park,drechsQneu,malek},
represent a family of quantum magnets for which the 1D $J_1$-$J_2$ Heisenberg model
is a good starting point for a theoretical description.
The above listed compounds are quasi-1D frustrated 
spin-$1/2$ magnets 
with a ferromagnetic (FM) nearest-neighbor (NN) in-chain coupling $J_1<0$ 
and an antiferromagnetic (AFM) next-nearest-neighbor (NNN)
in-chain coupling
$J_2>0$.

\vskip1mm
{\it The model:}
The Hamiltonian of the 1D $J_1$-$J_2$ Heisenberg ferromagnet is given by
\begin{equation}  \label{Ham}
  H=J_1\sum_{\langle i,j\rangle}{\bf S}_i{\bf S}_j+J_2\sum_{[ i,j]}{\bf S}_i{\bf
S}_j \; ,
\end{equation}
where the first sum runs over the NN bonds and  the second sum  over the NNN bonds. 
Henceforth we set $J_1=-1$.
For the model (\ref{Ham}) a quantum critical point at $J_2=0.25$ exists where 
the 
FM ground state (GS)
gives way for a
singlet GS with spiral correlations for $J_2 >
0.25$ \cite{bader,Hamada,krivnov}. 
For most of the edge-shared chain
cuprates $J_2$ is large enough to realize such a spiral GS. 
However, several materials considered as model systems for 1D
spin-1/2 ferromagnets, 
such as 
TMCuC[(CH$_3$)$_4$NCuCl$_3$] \cite{TMCuC}
and p-NPNN (C$_{13}$H$_{16}$N$_3$O$_4$) \cite{NPNN}, might have also a weak frustrating  
NNN interaction $J_2 <
0.25$. Moreover, recent studies \cite{malek} lead to the conclusion that Li$_2$CuO$_2$
is a quasi-1D spin-$1/2$ system with a dominant FM
$J_1$  and
weak frustrating AFM
$J_2 \approx 0.2|J_1| $.
Here
we focus on the parameter region $J_2 \le 0.25$, i.e.\   
the GS is ferromagnetic. 
Only at $J_2=0.25$ the FM GS multiplet  is 
degenerate with a spiral singlet GS \cite{bader,Hamada,krivnov}.  
On the other hand, the frustrating 
$J_2$ influences the low-energy excitations, in particular, if $J_2$ is close
to the quantum critical point. Hence, the frustration may have a strong
effect on the low-$T$ thermodynamics.
We mention  that previous studies \cite{tmrg,heidrich06} of the thermodynamics of the 
1D $J_1$-$J_2$
model with FM $J_1$
did not consider values of $J_2$ 
near
the quantum
critical point $J_2 \lesssim 0.25$.

\vskip1mm
{\it Results:}
To study the thermodynamic properties
we use the full exact diagonalization (ED) of finite
rings of up to $N=22$ lattice sites, 
complemented  by data obtained by a  
spin-rotation-invariant second-order Green-function method
(RGM) \cite{kondo,SSI94,rgm_new,hartel2008}. 
Note that by contrast to ED the RGM is limited to values $J_2 \le 0.2$
\cite{hartel2008} but yields results for $N\to \infty $, that allows the
calculation of the correlation length by the RGM. Here we will
present data for the spin-spin correlation functions  $\langle{\bf S}_0{\bf
S}_n\rangle$,
the uniform static spin susceptibility $\chi$ and the specific heat $C_v$.
For the discussion of the correlation length of the model (\ref{Ham}),
see Ref.~\cite{hartel2008}.  
For the unfrustrated model we will compare our results  with
available 
Bethe-ansatz data \cite{yamada1} and  transfer-matrix
renormalization group (TMRG) results \cite{tmrg}.
\begin{figure}[b]
\begin{minipage}{16pc}
\includegraphics[width=16pc]{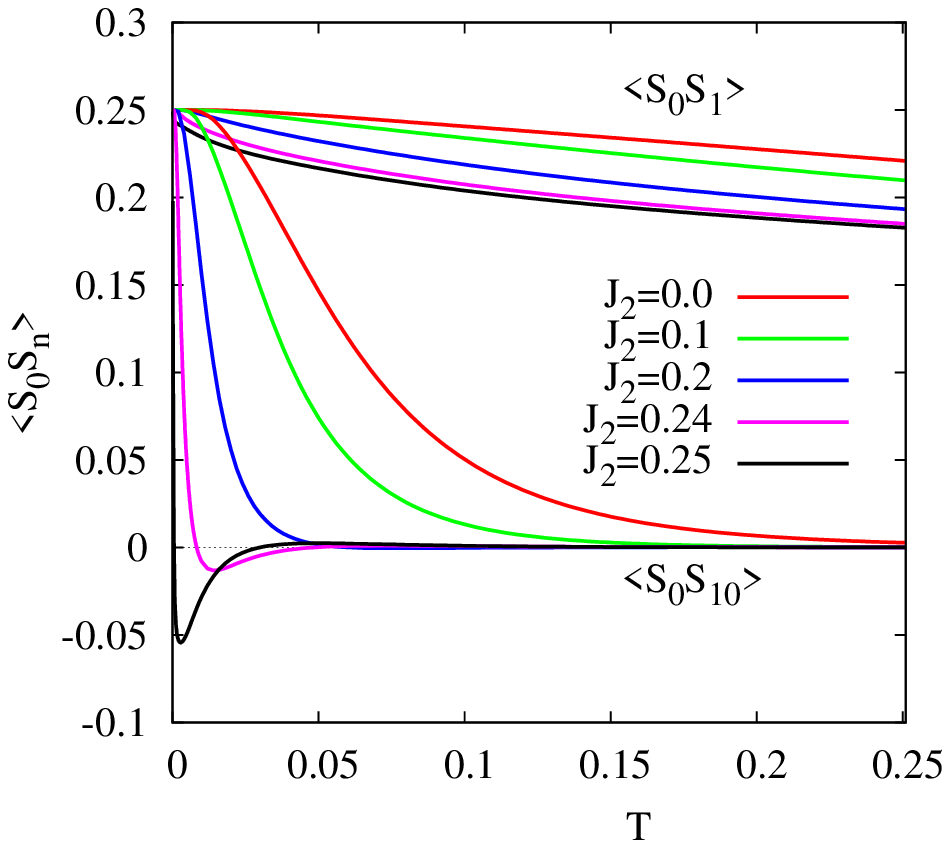}
\caption{\label{fig1}Spin correlation function $\langle{\bf S}_0{\bf
S}_1\rangle$ (NN) and $\langle{\bf S}_0{\bf
S}_{10}\rangle$ calculated by ED for $N=20$ sites.
}
\end{minipage}\hspace{2pc}%
\begin{minipage}{20pc}
\includegraphics[width=18pc]{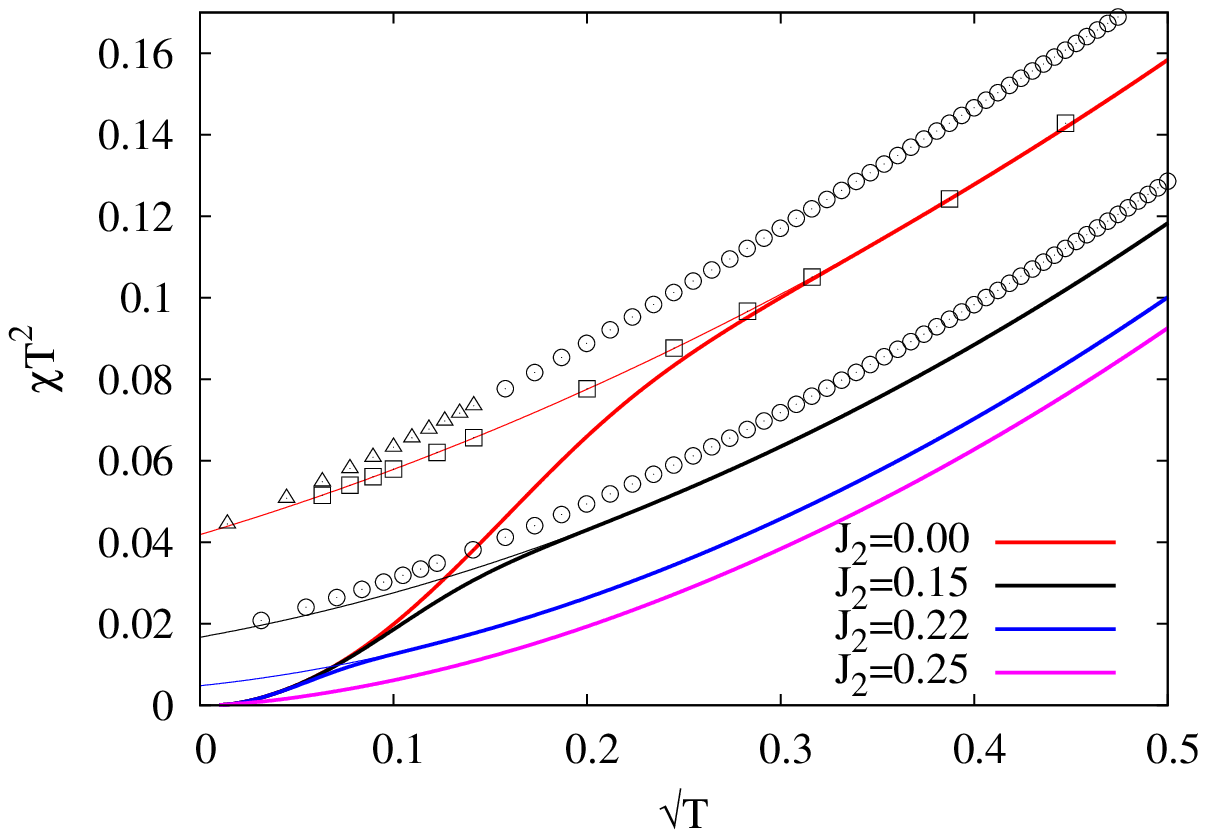}
\caption{\label{chi_T2} $\chi T^2$ versus $\sqrt{T}$ calculated  
by ED for $N=22$ (thick lines -- calculated data, thin lines --
extrapolation to $T \to
0$, see Eq.~(\ref{y_0}) and text)  and RGM (circles) as well as  
Bethe-ansatz data
(squares) for $J_2=0$ from Ref.~\cite{yamada1}.
The inset shows the coefficient $y_0 = \lim_{T \to 0} \chi T^2$ 
versus $J_2$.
}
\end{minipage} 
\end{figure}

The temperature dependence of the spin correlation functions 
$\langle{\bf S}_0{\bf S}_n\rangle$ 
is
shown for $n=1$ (NN) and $n=10$ for
various $J_2$ in Fig.~\ref{fig1}.
With increasing 
frustration the correlation functions decrease, where 
the further-distant 
correlators decay much stronger than the NN correlator. Near and at the quantum
critical point the large-distant correlator 
$\langle{\bf S}_0{\bf S}_{10}\rangle$ vanishes 
already at $T \gtrsim 0.05$. 
Interestingly, for $J_2=0.2$, $0.24$, and $0.25$
the correlator $\langle{\bf S}_0{\bf S}_{10}\rangle$ changes the sign
and goes through a minimum.
This behavior is not affected by finite-size effects, e.g., the 
correlators  $\langle{\bf S}_0{\bf S}_{8}\rangle$ for $N=16, 20$ and
$\langle{\bf S}_0{\bf
S}_{6}\rangle$ for $N=12, 16, 20$ also change the sign
and go through a minimum for  $J_2=0.2$, $0.24$, and $0.25$.

Next we discuss the low-temperature properties of the 
susceptibility 
$\chi = \lim_{h \to 0} d\langle S_z \rangle /dh$.
Due to the FM
GS $\chi$
diverges at $T \to 0$. Using Bethe-ansatz for $J_2=0$ 
the critical behavior has been determined as 
$\chi \propto  T^{-2}$ \cite{yamada1}. 
Using the RGM, recently it
has been confirmed  that the critical indices for the susceptibility and
the correlation length, $\gamma=2$ and $\nu=1$, respectively, 
are not changed by frustration for $J_2 < 0.25$. 
However, at the quantum critical point $J_2=0.25$ a change of the 
low-temperature
physics is expected \cite{sachdev}.   
To study that question 
we  consider 
the expansion
\begin{equation}\label{y_0} 
 \; \chi T^2 = y_0 + y_1 \sqrt{T} + y_2 T+  
{\cal O}(T^{3/2})\quad , 
\end{equation}
related  to the existence of the
FM
critical
point at $T=0$. It 
has been derived for $J_2=0$ in   
Ref.~\cite{yamada1}. 
For the frustrated system (\ref{Ham}) 
the coefficients $y_0$, $y_1$, and $y_2$ depend on $J_2$.
In Fig.~\ref{chi_T2} we plot
$\chi T^2$ versus $\sqrt{T}$.
We find a good agreement of the ED data for $\chi T^2$ with Bethe-ansatz
results
down to quite low temperature $T$. The RGM results for $\chi T^2$
deviate slightly from the
Bethe-ansatz results
for finite $T$, but approach the 
Bethe-ansatz 
data for $T \to
0$, see also Ref.~\cite{SSI94}.  
The behavior of the leading coefficient $y_0$ and the next-order
coefficient $y_1$   
can be extracted from the results for $\chi T^2$ 
by fitting them to Eq.~(\ref{y_0}).  
For the RGM we use  
data points up to a cut-off temperature $T=T_{cut}=0.005$.
To deal with finite-size effects in the ED data at very low
$T$, 
we use the specific heat per site $C_v(T)$, see below,  
to determine that temperature $T_{ED}$ down 
to which the
first four digits of $C_v(T)$ for 
$N=20$ and $N=22$ coincide.
Then we fit the ED data 
in the interval $T_{ED}\le T \le T_{ED} + T_{cut}$ 
to Eq.~(\ref{y_0}). Note that $T_{ED}$ becomes smaller for
increasing $J_2$, we find e.g.,
$T_{ED}=0.22, 0.13, 0.09, 0.04, 0.03, 0.02$ at $J_2=0.0, 0.1, 0.15, 0.2,
0.24, 0.25$, respectively.
For $J_2=0$ we found 
$y_0=1/24$ 
($y_0=0.0418$) for the RGM (ED), which
agrees with
the Bethe-ansatz results of Ref.~\cite{yamada1}.
[Note the different definitions of $\chi$ in our paper
and in Ref.~\cite{yamada1}.]
Including frustration $J_2>0$ we observe  a linear
 decrease of $y_0$ with $J_2$
down to zero at $J_2=0.25$ given by 
\begin{equation}
\;\; y_0 =(1-4J_2)/24 
\; ,
\end{equation}
cf. the inset of  Fig.~\ref{chi_T2}.
The vanishing of $y_0$ at $J_2=0.25$ 
indicates the change of the
low-$T$ behavior of the physical quantities at the quantum critical
point \cite{sachdev}.  
Indeed, 
a polynomial fit according to $y_1 = a_y + b_y J_2+ c_y J_2^2$  
yields the finite value 
$y_1 \approx
0.05 \; (0.04)$ for RGM (ED).
Hence, our data provide evidence  for a change of the low-$T$
behavior of $\chi$  from $\chi \propto T^{-2}$ 
at $J_2 < 0.25$ to $\chi \propto T^{-3/2}$ 
at the quantum critical point
$J_2=0.25$.      
For a
a similar discussion of
the correlation length $\xi$,  
see
Ref.~\cite{hartel2008}, 
where it was found that 
the low-$T$
behavior of $\xi$  changes from 
$\xi \propto
T^{-1}$ at $J_2 < 0.25$ to $\xi \propto
T^{-1/2}$ at  $J_2=0.25$.
\begin{figure}[t]
\begin{minipage}{18pc}
\includegraphics[width=18pc]{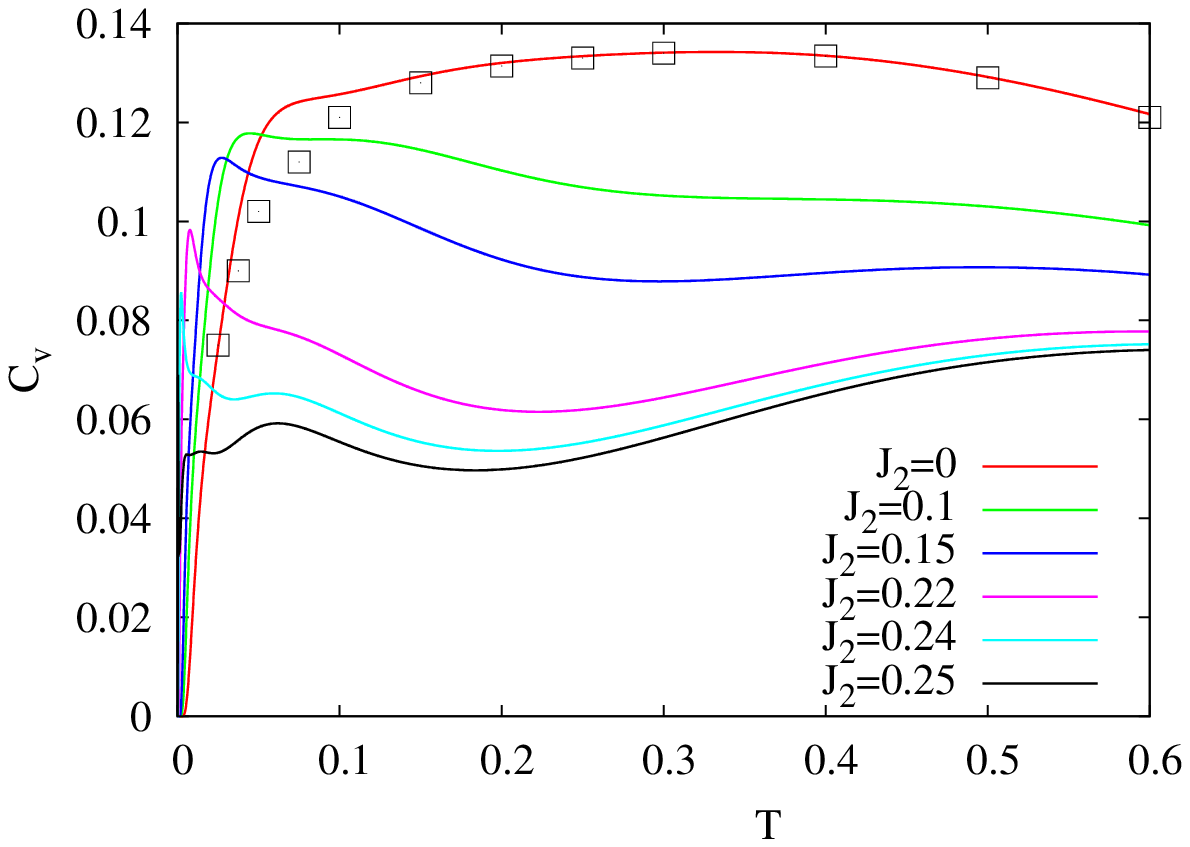}
\caption{\label{C_T_versch_J2}
ED data for the specific heat for $N=22$.
For comparison we show  
TMRG data
(squares) from Ref. \cite{tmrg} for $J_2=0$.}
\end{minipage}\hspace{2pc}%
\begin{minipage}{18pc}
\includegraphics[width=18pc]{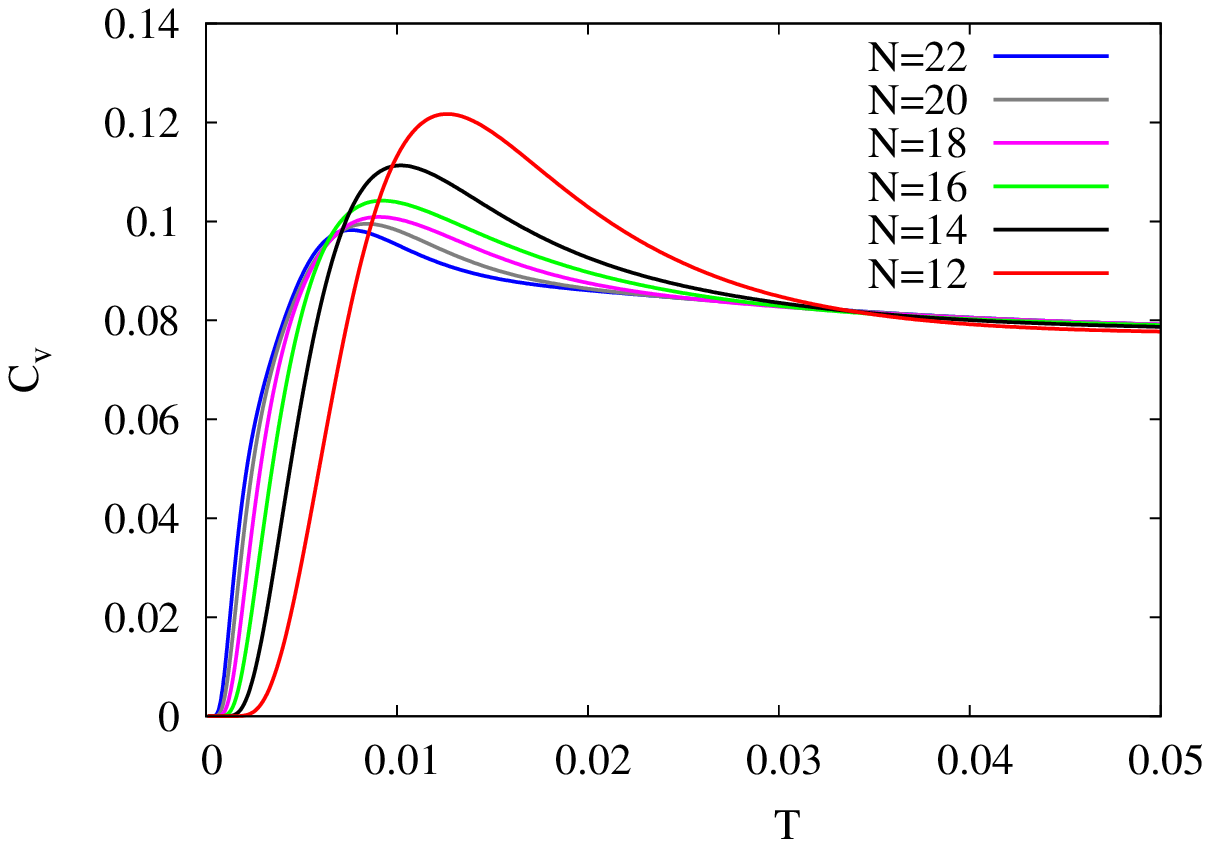}
\caption{\label{c-h0-verschN} Finite-size dependence of the 
specific heat for $J_2=0.22$.
The inset shows $C^{p}_{\infty}$ and $T^{p}_\infty$ versus $J_2$, see text. 
}
\end{minipage} 
\end{figure}

In Fig.~\ref{C_T_versch_J2} we present ED results for the specific heat $C_v$.
For $J_2=0$ we found a broad maximum 
at $T\approx 0.332$ and a steep decay to zero starting
at about 
$T = 0.05 $ 
in accord with the TMRG
\cite{tmrg}.
For  $J_2 \gtrsim 0.125$ 
the specific heat exhibits a minimum located at around $T=0.2$, and two
maxima, namely a  high-$T$ maximum at
around  $T= 0.6 $ and an additional low-$T$
maximum at $T < 0.1$. 
If $J_2$ approaches  $J_2=0.25$, 
a further quite sharp
peak at very low $T$
appears, that is, however, strongly size dependent, see
Fig.~ \ref{c-h0-verschN}.   
From Fig.~\ref{c-h0-verschN} it is obvious that
the extra low-$T$
finite-size peak behaves
monotonously with $N$.
Hence, we have performed a finite-size extrapolation to $N\to \infty$ 
of the height $C^{p}$ and the position
$T^{p}$ of the peak in $C_v(T)$ 
using the  formula 
$a(N)=a_\infty + a_1/N^2 + a_2/N^4$.
The extrapolated values $C^{p}_{\infty}$ and $T^{p}_\infty$ are shown in the
insets of Fig.~\ref{c-h0-verschN}. 
Obviously, $C^{p}_\infty >0$
even near the quantum critical
point $J_1=0.25$, where $C^{p}_\infty \approx 0.05$.  
On the other hand,
$T^{p}_\infty$ decreases with $J_2$ and becomes very small near  
$J_2= 0.25$.
This behavior suggests that a characteristic steep decay of
$C_v(T)$ down to zero starts at very low $T$
when approaching $J_2=0.25$.

\vskip1mm
{\it Summary:}
We discussed the thermodynamics of frustrated FM
spin-1/2 $J_1$-$J_2$ Heisenberg chains and found 
as prominent features 
(i) a change of the low-$T$
critical behavior at
the quantum critical point $J_2=|J_1|/4$, (ii) and an 
additional low-$T$ 
maximum in the specific heat for $|J_1|/4 > J_2 \gtrsim |J_1|/8$.  

\ack 
This work was supported by the DFG (projects RI615/16-1 and DR269/3-1).
One of us (S.-L. D.) is indebted to V.Ya.Krivnov for useful discussions.

\end{document}